\documentclass[a4paper,11pt]{article}
\pdfoutput=1 

\usepackage{arxiv} 

\usepackage[T1]{fontenc} 

\usepackage{cancel}
\usepackage[table]{xcolor}
\usepackage{slashed}
\usepackage{amsmath}  
\usepackage{amssymb}  
\usepackage{multirow}
\usepackage{mathtools}
\usepackage{mathrsfs}
\newcommand{\fsl}[1]{{\ooalign{\(#1\)\cr\hidewidth\(/\)\hidewidth\cr}}}
\newcommand*{\Scale}[2][4]{\scalebox{#1}{\ensuremath{#2}}}

\usepackage{ragged2e}

\makeatletter
\newcommand{\raisemath}[1]{\mathpalette{\raisem@th{#1}}}
\newcommand{\raisem@th}[3]{\raisebox{#1}{$#2#3$}}
\makeatother

\newcommand{\hodge}{{\star}}

\title{Generic axion Maxwell equations: path integral approach}

\author[a]{Anton V. Sokolov,}
\author[\,b]{Andreas Ringwald}

\affiliation[a]{Rudolf Peierls Centre for Theoretical Physics, University of Oxford, Parks Road, Oxford OX1 3PU, \\
United Kingdom}
\affiliation[b]{Deutsches Elektronen-Synchrotron DESY, Notkestr. 85, 22607 Hamburg, Germany}

\emailAdd{anton.sokolov@physics.ox.ac.uk}
\emailAdd{andreas.ringwald@desy.de}

\abstract{Using the path integral approach, we derive the low energy interactions between axions and electromagnetic fields that arise in models with heavy dyons charged under a spontaneously broken global axial $U(1)$ symmetry. Hence, we obtain generic axion-Maxwell equations relevant for experimental searches. We find that the structure of the axion Maxwell equations could be significantly different compared to what is normally assumed in the literature, as the derived equations feature new axion-dependent terms including CP-violating ones. The new terms can reconcile the Peccei-Quinn solution to the strong CP problem with astrophysical axion hints, as well as give unique signatures in light-shining-through-wall and haloscope experiments. Moreover, via the latter signatures, these experiments can indirectly probe the existence of heavy dyons.}


\begin{document}
    \begin{flushright}
		DESY-23-031\\
    \end{flushright}
    \maketitle
    \flushbottom

\section{Introduction}

Axions as hypothetical new particles are well motivated candidates for physics beyond the Standard Model (SM). In particular, axions provide a straightforward solution to the strong CP problem~\cite{Peccei:1977hh, Peccei:1977ur, Weinberg:1977ma, Wilczek:1977pj} and give a simple explanation for the inferred dark matter abundance and properties~\cite{Preskill:1982cy,Abbott:1982af,Dine:1982ah,Arias:2012az}. An advantage of the axion hypothesis is that it can be relatively easily probed by various experiments, most of which aim to detect the coupling of axions to the electromagnetic field. Due to the fact that most experiments focus on the electromagnetic coupling, it is essential to understand how the Maxwell equations change in the presence of axions, i.e. to derive the most general form of the so-called axion Maxwell equations. The widely accepted form of these equations is~\cite{Sikivie:1983ip}:
\begin{eqnarray}
    \label{axmax1}
	&&\pmb{\nabla}\! \times \! \mathbf{B}_a - \dot{\mathbf{E}}_a = -g_{a\gamma \gamma} \left( \mathbf{E}_0\! \times \! \pmb{\nabla} a - \dot{a} \mathbf{B}_0 \right) \, , \\[3pt]\label{axmax2}
	&&\pmb{\nabla}\! \times \! \mathbf{E}_a + \dot{\mathbf{B}}_a = 0 \, , \\[3pt]\label{axmax3}
	&& \pmb{\nabla}\! \cdot \! \mathbf{B}_a = 0 \, , \\[3pt]\label{axmax4}
	&& \pmb{\nabla}\! \cdot \! \mathbf{E}_a = -g_{a\gamma \gamma}\, \mathbf{B}_0\! \cdot \! \pmb{\nabla} a \, ,
\end{eqnarray}
where $\mathbf{E}_a$ and $\mathbf{B}_a$ are axion-induced electric and magnetic fields, while $\mathbf{E}_0$ and $\mathbf{B}_0$ are background electric and magnetic fields created in the detector. 

It has recently been shown~\cite{Sokolov:2022fvs} by the authors of this article that the axion Maxwell equations~\eqref{axmax1}--\eqref{axmax4} represent the special case of a more general construction which involves three axion coupling parameters $g_{a\mbox{\tiny{EE}}}$, $g_{a\mbox{\tiny{MM}}}$ and $g_{a\mbox{\tiny{EM}}}$ instead of the only one $g_{a\gamma \gamma}$ normally considered:
\begin{eqnarray}\label{genaxmax10}
	&&\pmb{\nabla}\! \times \! \mathbf{B}_a - \dot{\mathbf{E}}_a = -g_{a\mbox{\tiny{EE}}} \left( \mathbf{E}_0\! \times \! \pmb{\nabla} a - \dot{a} \mathbf{B}_0 \right) - g_{a\mbox{\tiny{EM}}} \left( \mathbf{B}_0\! \times \! \pmb{\nabla} a + \dot{a} \mathbf{E}_0 \right) \, , \\[3pt]\label{genaxmax20}
	&&\pmb{\nabla}\! \times \! \mathbf{E}_a + \dot{\mathbf{B}}_a =  g_{a\mbox{\tiny{MM}}} \left( \mathbf{B}_0\! \times \! \pmb{\nabla} a + \dot{a} \mathbf{E}_0 \right) + g_{a\mbox{\tiny{EM}}} \left( \mathbf{E}_0\! \times \! \pmb{\nabla} a - \dot{a} \mathbf{B}_0 \right) \, , \\[3pt]\label{genaxmax30}
	&& \pmb{\nabla}\! \cdot \! \mathbf{B}_a = g_{a\mbox{\tiny{MM}}}\, \mathbf{E}_0\! \cdot \! \pmb{\nabla} a - g_{a\mbox{\tiny{EM}}}\, \mathbf{B}_0\! \cdot \! \pmb{\nabla} a \, , \\[3pt]\label{genaxmax40}
	&& \pmb{\nabla}\! \cdot \! \mathbf{E}_a = - g_{a\mbox{\tiny{EE}}}\, \mathbf{B}_0\! \cdot \! \pmb{\nabla} a + g_{a\mbox{\tiny{EM}}}\, \mathbf{E}_0\! \cdot \! \pmb{\nabla} a \, .
\end{eqnarray}
It was found that such new electromagnetic couplings of axions arise in KSVZ-like models~\cite{Kim:1979if, Shifman:1979if} where the new heavy particles charged under the Peccei-Quinn (PQ) $U(1)_{\rm PQ}$ group carry magnetic charges~\cite{Sokolov:2021ydn, Sokolov:2021eaz}. As this construction generalizes the conventional axion models to the case where there exist heavy dyons, we dub these models as dyon-philic. The latter models are well motivated from the theoretical viewpoint as there is no reason to expect a new heavy particle to carry only an electric, but no magnetic  charge~\cite{Dirac:1931kp}; moreover, the existence of heavy magnetically charged particles seems to be a necessity stemming from the quantization of the electric charge observed in nature~\cite{Banks:2010zn, Harlow:2018tng, Polchinski:2003bq}. 

Note that by definition, background fields $\mathbf{E}_0$ and $\mathbf{B}_0$ satisfy free Maxwell equations. It is essential that these fields are treated as external. Indeed, close to the sources, which have to be treated as point-like in a classical theory with both electric and magnetic charges~\cite{Dirac:1948um, Brandt:1978wc}, the latter fields diverge and the classical weak field approximation used in Eqs.~\eqref{genaxmax10}-\eqref{genaxmax40} is no longer valid. 

While in Ref.~\cite{Sokolov:2022fvs} we focused primarily on explaining the loophole in the previous theoretical works on the axion-photon coupling and investigated the new electromagnetic couplings of axions in the Effective Field Theory (EFT) approach, in this article we would like to give a detailed derivation of the general axion Maxwell equations~\eqref{genaxmax10}-\eqref{genaxmax40} in the path integral framework. The path integral approach is very instructive when dealing with the Quantum Field Theory (QFT) that includes both electric and magnetic charges: indeed, the only exhaustive proof of the Lorentz-invariance of such QFT was obtained in the path integral framework~\cite{Brandt:1977be, Brandt:1978wc}. Moreover, in the path integral approach, it is easier to understand how the peculiarities of the corresponding classical theory, such as Dirac vetos, originate and why they do not signal any inconsistencies.

This article is structured as follows: in sec.~\ref{qftmagn}, we review the path integral formulation of the QFT describing the interactions between electric and magnetic charges as well as discuss the general distinctive features of this QFT; in sec.~\ref{sec3}, we introduce the dyon-philic axion models in the path integral approach, transform the integral over the heavy dyon field into the integral over the dyon proper time parameter, and calculate the effective low energy Lagrangian that describes the interactions of dyon-philic axions with the electromagnetic field by performing an exact non-perturbative calculation; in sec.~\ref{sec4}, we derive the general axion Maxwell equations~\eqref{genaxmax10}--\eqref{genaxmax40} and briefly discuss their experimental implications; finally, in sec.~\ref{seclast}, we conclude.

\section{QFT with magnetic charges and its path integral formulation}\label{qftmagn}

There exist several equivalent formulations of the QFT with magnetic charges, see e.g. Refs.~\cite{Schwinger:1966nj, PhysRevD.3.880}, and the exhaustive  review~\cite{Blagojevic:1985sh}. All these formulations necessarily share a common feature which drastically distinguishes them from the theory of Quantum Electrodynamics (QED). This feature is non-locality\footnote{We use the term non-locality in the sense of quantum non-locality, i.e. entanglement. Microcausality is preserved.}: any QFT with both electric ($e_i$) and magnetic ($g_j$) charges is actually a theory of the interaction of two-particle irreducible states~\cite{Zwanziger:1972sx, Csaki:2020inw, Csaki:2020yei}. Indeed, an asymptotic state of two particles $(i,j)$ for which $e_i g_j - e_j g_i \neq 0$ constitutes an entangled ``pairwise" state. 
Each such state is characterized by a pairwise helicity variable which corresponds to the asymptotically non-vanishing angular momentum of the electromagnetic field. Note that the Dirac-Schwinger-Zwanziger (DSZ) quantization condition follows naturally from the quantization of this angular momentum. Asymptotic irreducible two-particle states obviously violate the cluster decomposition principle, which is another way to understand why any QFT with magnetic charges is fundamentally different from QED, for which the cluster decomposition is a well established property (see e.g. Ref.~\cite{Nakanishi:1990qm}, pp.~252-254). While one can introduce magnetic monopoles in QED as external sources, to study the creation or annihilation of real and virtual monopoles one has to work in a substantially different framework.

The QFT of electric and magnetic charges which we choose to work with is Zwanziger theory~\cite{PhysRevD.3.880}. Zwanziger's formulation has the advantage of featuring a local Lagrangian and, as a consequence, of being the most extensively studied QFT with magnetic charges. The Lagrangian of the Zwanziger theory is:
\begin{eqnarray}\label{Zwanz}
	    &&\mathcal{L}_Z \; = \; \frac{1}{2} \left\lbrace \left[ n\! \cdot \! (\partial \wedge B) \right] \cdot [ n\! \cdot \! (\partial \wedge A)^d ] \; - \; \left[ n\! \cdot \! (\partial \wedge A) \right] \cdot [ n\! \cdot \! (\partial \wedge B)^d ] \; - \right. \nonumber \\
        &&\qquad \qquad \qquad \qquad \qquad \qquad \left. \left[ n\! \cdot \! (\partial \wedge A) \right]^2 \; - \; \left[ n\! \cdot \! (\partial \wedge B) \right]^2 \right\rbrace  \; - \; j_e \! \cdot \! A \; - \; j_m \! \cdot \! B \; ,
\end{eqnarray}
where $j_e$ and $j_m$ are electric and magnetic currents, respectively; $A_{\mu}$ and $B_{\mu}$ are four-potentials; $n_{\mu}$ is a fixed four-vector ($n^2 = 1$). We use the following simplified notations: $a \cdot b = a_{\mu} b^{\mu}$, $(a\wedge b)^{\mu \nu} = a^{\mu} b^{\nu} - a^{\nu} b^{\mu} , \,\, (a \cdot  G)^{\nu} = a_{\mu} G^{\mu \nu}$, and for any tensor $A_{\mu \nu}$ its Hodge dual is defined as ${A}^d_{\mu \nu} = \epsilon_{\mu \nu \lambda \rho} A^{\lambda \rho} / 2$, where $\epsilon_{0123} = 1$. Note that contrary to the case of QED, the Zwanziger theory features two four-potentials instead of one.  Still, the corresponding dynamical system is highly constrained, that is why one has only four phase space degrees of freedom describing the electromagnetic field, similar to QED~\cite{PhysRevD.3.880, Balachandran:1974nw}\footnote{Probably the simplest explanation of why there are two four-potentials, but the same number of degrees of freedom as in QED, is that these two four-potentials each satisfy a first-order differential equation in time, instead of the second-order differential equation satisfied by the only four-potential of QED. A system of $2N$ first-order equations has the same number of degrees of freedom as a system of $N$ second-order equations. The quantum state of the free electromagnetic field within the Zwanziger theory is equivalent to the similar state within QED, as it was explicitly shown in Ref.~\cite{Balachandran:1974nw}.}. The presence of the fixed four-vector $n_{\mu}$ in the Lagrangian is another very important feature of the Zwanziger theory. This feature is tightly connected to the non-locality property discussed in the previous paragraph, as the components of the field strength tensors $(\partial \wedge A)_{\mu \nu}$ and $(\partial \wedge B)_{\mu \nu}$ corresponding to the two four-potentials differ from the physical electric and magnetic fields by non-local $n_{\mu}$-dependent terms:
\begin{eqnarray}\label{elect}
	\partial \wedge A = F + (n\! \cdot \! \partial)^{-1} (n\wedge j_m)^d \, , \\ \label{magnet}
	\partial \wedge B = F^d - (n\! \cdot \! \partial)^{-1} (n\wedge j_e)^d \, , 
\end{eqnarray}
where $F_{\mu \nu}$ is the electromagnetic field strength tensor and $(n\! \cdot \! \partial)^{-1}$ is the integral operator satisfying $n \cdot \partial \left( n \! \cdot \! \partial \right)^{-1} \! \left(\vec{x} \right) = \delta \! \left( \vec{x} \right) $. The major advantage of the two four-potentials of the Zwanziger theory is their regularity everywhere in space-time, i.e. $ \forall x_{\mu} : \left[ \partial_{\rho} , \partial_{\nu}  \right] A_{\lambda} (x_{\mu}) = \left[ \partial_{\rho} , \partial_{\nu}  \right] B_{\lambda} (x_{\mu}) = 0$. This property is satisfied only due to the field decompositions~\eqref{elect} and~\eqref{magnet}.

The dependence of the Lagrangian~\eqref{Zwanz} on the fixed four-vector $n_{\mu}$ means that a special attention should be paid to the Lorentz-invariance of the theory. One can check straight away that the classical equations of motion for the electromagnetic field do not depend on $n_{\mu}$ and are thus Lorentz-invariant by simply varying the Lagrangian~\eqref{Zwanz} with respect to the four-potentials $A_{\mu}$ and $B_{\mu}$ and using Eqs.~\eqref{elect} and~\eqref{magnet}. One of course obtains the classical Maxwell equations this way. It is a bit more difficult to see that the classical equations of motion for the charged particles do not depend on $n_{\mu}$. Writing the currents in terms of point-particle trajectories:
\begin{eqnarray}\label{je}
	j_e^{\nu} (x) = \sum_i e_i \int \delta^4 (x-x_i(\tau_i))\, dx_i^{\nu} \, , \\ \label{jm}
	j_m^{\nu} (x) = \sum_i g_i \int \delta^4 (x-x_i(\tau_i))\, dx_i^{\nu} \, , 
\end{eqnarray}
and varying with respect to these trajectories, one obtains:
\begin{eqnarray}\label{force}
	\frac{d}{d\tau_i}\! \left( \frac{m_i u_i}{(u_i^2)^{1/2}} \right) = \!\! && \left( e_i F(x_i) + g_i F^d(x_i) \, \right) \! \cdot \! u_i  \nonumber \\
	&& - \sum_j (e_i g_j - g_i e_j) \, n\! \cdot \!\! \int (n\! \cdot \! \partial)^{-1} (x_i - x_j) \,\, (u_i \wedge u_j)^d\, d\tau_j \, .
\end{eqnarray}
The second term on the right-hand side seems to spoil both the Lorentz-invariance of the classical theory and the agreement with the conventional expression for the Lorentz force. However, it is easy to see that this term does not contribute to the dynamics: the support of the kernel $(n\! \cdot \! \partial)^{-1} (x_i - x_j)$ is given by the condition $ \vec{x}_i (\tau) - \vec{x}_j (\tau) = \vec{n} s $, which has more equations than free parameters and is thus satisfied only for exceptional trajectories. As such exceptional trajectories form a measure zero subset of all possible trajectories, one can safely omit them in the variational procedure: indeed, the latter procedure originates from calculating the path integral over all the trajectories where no measure zero subset can contribute. One can also rigorously exclude the exceptional trajectories from consideration within the classical theory itself by redefining the action functional as suggested in Ref.~\cite{Brandt:1976hk}.

We explained why the classical equations of the Zwanziger theory are $n_{\mu}$-independent. It remains to be shown that the full quantum theory shares this property. The respective proof was given in Refs.~\cite{Brandt:1977be, Brandt:1978wc}. We will briefly return to their arguments after presenting the path integral formulation of the theory. Before this, let us note that at each finite order of the formal perturbation theory applied to the Zwanziger theory, the resulting approximation is \textit{not} $n_{\mu}$-independent and is therefore ill-defined. This means that the theory is essentially non-perturbative. A very important point is that this non-perturbativity is associated to the $n_{\mu}$-dependence and thus to the non-locality feature described earlier, but not necessarily to the presence of a large expansion parameter. Indeed, the effective expansion parameter can be made small for some particular processes~\cite{Drukier:1981fq, Graf:1991xe, DeRujula:1994nf}, however this does not justify applying perturbation theory: the latter is still ill-defined and can give wrong results as explained in Ref.~\cite{Ignatiev:1997pm}.
 
Zwanziger theory was originally quantized using the canonical formalism, either by adding a special gauge-fixing term~\cite{PhysRevD.3.880} or by invoking the full Dirac's method for the quantization of constrained systems~\cite{Balachandran:1974nw}. In this work, we are however interested in the path integral approach. A thorough path integral formulation of the Zwanziger theory was given in Ref.~\cite{Senjanovic:1976br}, developed further in Refs.~\cite{Brandt:1977be, Brandt:1978wc} and refined by using lattice regularization in Ref.~\cite{Calucci:1982wy}. Choosing the gauge-fixing functions to be
\begin{eqnarray}
    &&G_1(\alpha) = n\!\cdot \! A - \alpha \, , \\
    &&G_2 (\beta) = n\! \cdot \! B - \beta \, ,
\end{eqnarray}
one obtains the following generating functional of the Green's functions of the theory:
\begin{eqnarray}
\mathcal{Z}(\tilde{a}_{\mu},\tilde{b}_{\mu}) \; = \;  \mathcal{N} \int \prod_{\mu} \mathcal{D} A_{\mu} \mathcal{D} B_{\mu} \prod_{x} \delta \! \left( n\!\cdot \! A - \alpha \right) \delta \! \left( n\!\cdot \! B - \beta \right) \exp \left\lbrace i \int d^4 x \, \left( \mathcal{L}_Z + j_e \!\cdot \! \tilde{a} + j_m \!\cdot \! \tilde{b} \right) \right\rbrace  \, , \quad \;
\end{eqnarray}
where $\mathcal{L}_Z$ is the Zwanziger Lagrangian~\eqref{Zwanz}; $\mathcal{N}$ is the normalization factor including the Faddeev-Popov determinant~\cite{Faddeev:1967fc} and the determinant associated to the second-class constraints~\cite{Senjanovic:1976br}, both of which are independent of the fields in this case; $\tilde{a}_{\mu}$ and $\tilde{b}_{\mu}$ are arbitrary functions. Note that we omitted the obvious part of the generating functional containing only charged matter fields and the functional integrations over them. Integrating over the parameters $\alpha$ and $\beta$ with suitable weights~\cite{Brandt:1977fa}, one can as usual trade the gauge-fixing conditions for the gauge-fixing terms in the Lagrangian:
\begin{eqnarray}\label{genfunc}
    \mathcal{Z}(\tilde{a}_{\mu},\tilde{b}_{\mu}) \; = \;  \mathcal{N} \int \prod_{\mu} \mathcal{D} A_{\mu} \mathcal{D} B_{\mu} \exp \left\lbrace i \int d^4 x \, \left( \mathcal{L}_Z +\mathcal{L}_G + j_e \!\cdot \! \tilde{a} + j_m \!\cdot \! \tilde{b} \right) \right\rbrace  \, ,
\end{eqnarray}
where
\begin{equation}\label{gaugefix}
	\mathcal{L}_{G} \; = \; \frac{1}{2} \left\lbrace \left[ \partial \left( n\! \cdot \! A \right) \right]^2 + \left[ \partial \left( n\!\cdot \! B \right) \right]^2  \right\rbrace \, .    
\end{equation}

The known proof of the Lorentz-invariance of the Zwanziger QFT~\cite{Brandt:1977be, Brandt:1978wc} relies essentially on the path integral representation of the theory. Let us briefly explain the main ideas behind this proof. First of all, to establish the Lorentz-invariance of the theory, it is sufficient to show the Lorentz-invariance of the generating functional~\eqref{genfunc}. Second, it is easy to notice that the Lorentz-invariance in this case is equivalent to $n_{\mu}$-independence. Finally and most importantly, one has to remember that the functional integrals over the charged matter fields can be represented as series involving integrals over point-particle trajectories~\cite{Feynman:1948ur, Feynman:1950ir, Schwinger:1951nm, Brandt:1978wc}. It then turns out that the only $n_{\mu}$-dependence remains in the interactions between particles (real or virtual) of a different electric-magnetic type  and that the $n_{\mu}$-dependent part basically counts the number of times the trajectory of one particle intersects some oriented $n_{\mu}$-dependent three-surface associated to the trajectory of another particle, which is simply an integer but for some exceptional trajectories that form a measure zero subset and can therefore be omitted in the integral over all the trajectories. The part associated to the $n_{\mu}$-dependent integer does not contribute to the generating functional after imposing the DSZ quantization condition $e_ig_j - e_j g_i = 2\pi m, \, m \in \mathbb{Z}$ on the charges of all the possible $(i,j)$ pairs of dyons, since in this case the $n_{\mu}$-dependent contribution to the action is always equal to $2\pi k, \, k\in \mathbb{Z}$ which obviously does not contribute to the path integral~\eqref{genfunc}. 

Note that both in the classical case and in the full quantum relativistic case, the $n_{\mu}$-independence crucially depends on the point-particle representation of charged matter, as opposed to the usual continuum approximation in field theory where the distribution of charged matter is continuous. In fact, the classical field theory of magnetic charges, i.e. the theory where charges and currents are by definition continuously distributed in space, is always inconsistent, as the Jacobi identity for the gauge covariant derivatives is necessarily violated in this case~\cite{Dirac:1948um, Brandt:1978wc}. While the failure of the continuum approximation seems to be against the usual local-field-theoretic intuition
, one has to remember that one of the key features of the QFT with magnetic charges is its non-locality. As $n_{\mu}$-vector in the Zwanziger theory is responsible for capturing the non-locality, it is not surprising that the point-particle, as opposed to the continuous, distribution of charge is crucial for $n_{\mu}$-independence of the gauge-invariant observables. Finally, let us note that the same feature of non-locality invalidates the conventional decoupling principle applied to the theories with both electric and magnetic charges: a given heavy charged particle cannot be fully integrated out at the energy scales below its mass since it contributes a non-local angular momentum to the electromagnetic field, which is felt by other particles even in the deep infrared (IR), cf. Eqs.~\eqref{elect} and~\eqref{magnet}. This means that even if all the magnetically charged particles are very heavy, and their effect on the low energy interactions is indirect, the low energy theory describing the interactions of light electrically charged particles is \textit{not} given by a QED-like theory, but still by a theory the structure of which is similar to the QFT with magnetic charge. Simply put, in this case, we still need Zwanziger-like (or any equivalent) description of the electromagnetic field even at low energies.

\section{Electromagnetic interactions of dyon-philic axions}\label{sec3}

\subsection{Outline of the model}
Let us now consider the dyon-philic axion models~\cite{Sokolov:2021ydn, Sokolov:2021eaz, Sokolov:2022fvs} in the path integral framework of the previous section. In these models, similarly to the KSVZ axion model~\cite{Kim:1979if, Shifman:1979if}, one introduces at least one new heavy vector-like quark $\psi$ charged under the global $U(1)_{\rm PQ}$ symmetry, as well as the PQ complex scalar field $\Phi$ which gives mass to the new quark(s) in the phase where the $U(1)_{\rm PQ}$ is spontaneously broken due to a non-zero vacuum expectation value $\left\langle \Phi \right\rangle = v_a/\sqrt{2}$. A new heavy quark can in general be charged under the electromagnetic $U(1)_{\rm EM}$ subgroup of the gauge group of the Standard model. Moreover, there is no reason to assume that it carries only an electric, but no magnetic charge, given that the existence of heavy magnetically charged particles is currently regarded as a necessity for obtaining a consistent quantum gravity theory~\cite{Banks:2010zn, Harlow:2018tng, Polchinski:2003bq}. According to the discussion of the previous section, to describe all the effects of magnetically charged particles, one has to work in the framework of the QFT with magnetic charges. In particular, we chose to work with a particular realization of the QFT with magnetic charges given by the Zwanziger formalism quantized via the path integral methods outlined in the previous section. 

The part of the Lagrangian which includes interactions of the new heavy quark $\psi$ with the electromagnetic field and the PQ field $\Phi$ is:
\begin{equation}\label{1}
\mathcal{L}_{\psi , \Phi} \; = \; i\bar{\psi} \gamma^{\mu} D_{\mu} \psi + y \left( \Phi \, \bar{\psi}_L {\psi}_R + \text{h.c.}\, \right) - \lambda_{\Phi} \left( \left| \Phi \right|^{2} -\frac{v_a^2}{2} \right)^{\!\! 2},
\end{equation}
where $y$ and $\lambda_{\Phi}$ are some $O(1)$ constants and $D_{\mu} = \partial_{\mu} - ie_{\psi}A_{\mu} - ig_{\psi} B_{\mu}$ with $e_{\psi}$ and $g_{\psi}$ being the electric and magnetic charges of $\psi$, respectively. Let us decompose $\Phi = \left( v_a + \sigma + i a \right) \! / \sqrt{2}$, where $a$ is a pseudo Goldstone axion field. In the symmetry broke phase, the field $\sigma$ gets a mass $m_{\sigma} \sim v_a$ and decouples from the physics of low energy  processes, i.e. the processes for which the square of the center-of-mass energy $s \ll v_a^2$. Assuming $v_a$ is sufficiently large, which is indicated by experimental results and cosmology, the field $\sigma$ is then irrelevant for experiments. The light axion field $a$ on the contrary can be probed by low energy experiments and in this work we are interested in its electromagnetic interactions mediated by the field $\psi$. The relevant part of the Lagrangian in the symmetry broken phase can then be written as follows:
\begin{equation}\label{lpsi}
	\mathcal{L}_{\psi} \; = \; i\bar{\psi} \gamma^{\mu} D_{\mu} \psi + \frac{y v_a}{\sqrt{2}}\, \bar{\psi} \psi + \frac{i y}{\sqrt{2}}\, a\bar{\psi} \gamma_5 \psi \, .
\end{equation}

Using the equations~\eqref{genfunc} and~\eqref{gaugefix} from the previous section as well as the Lagrangian for the heavy field $\psi$ given by Eq.~\eqref{lpsi}, one can now write the generating functional of Green's functions of the theory in the symmetry broken phase:
\begin{eqnarray}\label{genfunc1}
        \mathcal{Z}(\tilde{a}_{\mu},\tilde{b}_{\mu}) \; = \;  \mathcal{N} \int \prod_{\alpha , \mu} \mathcal{D} \psi_{\alpha} \mathcal{D} \bar{\psi}_{\alpha}  \mathcal{D} A_{\mu} \mathcal{D} B_{\mu} \exp \left\lbrace i \int d^4 x \, \left( \mathcal{L}_Z + \mathcal{L}_G + \mathcal{L}_{\psi} + j_e \!\cdot \! \tilde{a} + j_m \!\cdot \! \tilde{b} \right) \right\rbrace  \, ,
\end{eqnarray}
where the parts associated to the axion field alone, i.e. kinetic energy and self-interaction, as well as the functional integration over the axion field, are omitted for sake of brevity; $j_e$ and $j_m$ are currents of any light charged particles that are used in our axion detectors. We omitted the functional integration over the corresponding light fermion fields as well as the kinetic terms associated to them. Practically, one has of course $j_m = 0$.

\subsection{Proper time representation of the heavy fermion path integral}
Using the generating functional~\eqref{genfunc1}, we would like to derive the low energy ($s\ll v_a^2$) interactions of axions with the electromagnetic field sourced by light charged particles. The calculation of the functional integral over $\psi$ yields:
\allowdisplaybreaks
\begin{eqnarray}\label{psitake}
    &&\!\!\!\!\!\!\!\!\!\! \mathcal{N} \int \prod_{\alpha} \mathcal{D} \psi_{\alpha} \mathcal{D} \bar{\psi}_{\alpha} \exp \left( i\! \int \! d^4 x\, \mathcal{L}_{\psi} \right) \; = \;  \exp \left\lbrace \text{Tr} \ln \left( i\fsl{D} + m +\frac{i y}{\sqrt{2}}\, a \gamma_5 + i \epsilon \right) \; - \right. \nonumber \\[6pt]
    &&\!\!\!\!\! \biggl. \text{Tr} \ln \biggl( i\fsl{\partial} + m  + i \epsilon \biggr) \biggr\rbrace  \; = \; \exp \biggl\lbrace \text{Tr} \ln \left( i\fsl{D} + m +\frac{i y}{\sqrt{2}}\, a \gamma_5 + i \epsilon \right) \; - \text{Tr} \ln \biggl( i\fsl{D} + m  + i \epsilon \biggr) \biggr\rbrace \times \nonumber \\[6pt]
    && \qquad \qquad \qquad \qquad \qquad \quad \qquad \qquad \qquad \exp \biggl\lbrace \text{Tr} \ln \biggl( i\fsl{D} + m + i \epsilon \biggr) \; - \text{Tr} \ln \biggl( i\fsl{\partial} + m  + i \epsilon \biggr) \biggr\rbrace \, .
\end{eqnarray}
where $m\equiv yv_a/\sqrt{2}$, the traces are over all the possible states,  and we normalized the integral by its value for the free fermion. We are interested in the first exponent on the right-hand side of Eq.~\eqref{psitake} since it describes interactions involving the axion field. We will transform the expression under this exponent by introducing the Schwinger proper time parameter. We choose the basis of states to be represented by position eigenstates, and sum over the spinor indices. Besides, as we are interested in low energy dynamics of axion, we omit the terms containing $\partial_{\mu} a$, as these terms are suppressed by $\omega_a/m$ and $|\mathbf{k}_a|/m$ compared to the others, where $\omega_a$ and $\mathbf{k}_a$ are the energy and momentum of the axion field, respectively. After we introduce the integration over the parameter $y$, the low energy approximation gives:
\begin{eqnarray}\label{ratio}
    && \text{Tr} \ln \left( i\fsl{D} + m +\frac{i y}{\sqrt{2}}\, a \gamma_5 + i \epsilon \right) \; - \text{Tr} \ln \biggl( i\fsl{D} + m  + i \epsilon \biggr) \; = \nonumber \\[6pt]
     &&\qquad \qquad \qquad \qquad \qquad \quad \qquad \qquad \qquad  \text{tr}_{\gamma} \int d^4 x \int\limits_{0}^{y/\sqrt{2}} d\tilde{y} \, \left\langle x \left| \, \frac{ia\gamma_5}{i\fsl{D} + m +i \tilde{y}\, a \gamma_5 + i \epsilon} \, \right| x \right\rangle \, ,
\end{eqnarray}
where $\text{tr}_{\gamma}$ denotes the trace over spinor indices, and the order of the Dirac matrices is unambiguous due to the trace operator. The resulting integral depends on the following dimensionful parameters: the axion field $a$, the electromagnetic field $\left[ D_{\mu}, D_{\nu} \right]$ and the mass $m$ of the heavy fermion. Taking into account gauge and Lorentz symmetries, it is clear that any terms describing the interaction of axions with the electromagnetic field are suppressed by some powers of $m$, and that the dominant terms are linear in the axion field $a$. This allows us to keep track only of the terms linear in $a$ in the low energy approximation. Working in the latter approximation, we rationalize the Dirac operator and introduce the integration over the Schwinger proper time parameter as follows:
\begin{eqnarray}\label{propt}
    &&\text{tr}_{\gamma} \int d^4 x \int\limits_{0}^{y/\sqrt{2}} d\tilde{y} \, \left\langle x \left| \, \frac{ia\gamma_5}{i\fsl{D} + m +i \tilde{y}\, a \gamma_5 + i \epsilon} \, \right| x \right\rangle \;\; = \;\; \frac{1}{2}\, \text{tr}_{\gamma} \int d^4 x \int\limits_{0}^{y/\sqrt{2}} d\tilde{y} \int\limits_{0}^{\infty} d\tau \, ia\gamma_5 \times \nonumber \\[6pt]
    &&  \!\!\!\!\!\!\!\! \qquad \qquad \; \langle x | \left( i \fsl{D} - m \right) e^{-i\tau (\Scale[0.7]{\fsl{D}}^2+m^2)/2} | x \rangle \;\; = \;\; -\text{tr}_{\gamma} \int d^4 x \; \frac{iym}{2\sqrt{2}}\, a \gamma_5 \int\limits_0^{\infty} d\tau \, \langle x | e^{-i\tau (\Scale[0.7]{\fsl{D}}^2+m^2)/2}  | x \rangle \, .
\end{eqnarray}

\subsection{Role of the non-local terms}
The proper time integral on the right-hand side of Eq.~\eqref{propt} can be calculated for certain electromagnetic field configurations, including a constant homogeneous field, using the Schwinger method~\cite{Schwinger:1951nm}. Since we are interested in dynamics at low energies, the scale of the variation of the electromagnetic field is negligible compared to the mass $m$ of the heavy fermion, so that the field can indeed be considered constant and homogeneous. However, due to the non-local terms in Eqs.~\eqref{elect} and~\eqref{magnet}, the constant homogeneous electromagnetic field is not automatically synonymous with constant homogeneous $\left[ D_{\mu}, D_{\nu} \right]$. To find out how to deal with the non-local terms, we will rewrite the functional integral Eq.~\eqref{genfunc1} in terms of the integrals over classical particle paths, as first suggested in Refs.~\cite{Brandt:1977be, Brandt:1978wc}. The matrix element in the integrand on the right-hand side of Eq.~\eqref{propt} corresponds to the following integral over trajectories:
\begin{eqnarray}\label{inttraja}
    &&\!\!\!\!\!\!\!\!\!\!\!\!\!\!\!\! \langle x | e^{-i\tau (\Scale[0.7]{\fsl{D}}^2+m^2)/2}  | x \rangle \; = \; e^{-i\tau m^2/2} \int\limits_{z(0)=x}^{z(\tau)=x} \mathcal{D}z (\tau)\, T \exp \Biggl\lbrace - i \int\limits_{0}^{\tau} d\tau' \left( \frac{\dot{z}^2}{2} + e_{\psi} A\!\cdot \! \dot{z} + g_{\psi} B\!\cdot \! \dot{z}\, + \right. \Biggr. \nonumber \\[6pt]
    &&\!\!\!\!\!\! \Biggl. \left. \frac{1}{4}\, \gamma^{\mu} \gamma^{\nu} \left[ D_{\mu},D_{\nu} \right] \right) \Biggr\rbrace \; = \; e^{-i\tau m^2/2} \int\limits_{z(0)=x}^{z(\tau)=x} \mathcal{D}z (\tau)\, \exp \Biggl\lbrace - i \int\limits_{0}^{\tau} d\tau' \left( \frac{\dot{z}^2}{2} + e_{\psi} A\!\cdot \! \dot{z} + g_{\psi} B\!\cdot \! \dot{z} \right) \Biggr\rbrace \times \nonumber \\[6pt]
    &&  \qquad \qquad \qquad \qquad \qquad \qquad \qquad \qquad \; \int d\Gamma (\tau) \, c(\tau) \otimes c^* (0) \exp \Biggl\lbrace -\frac{i}{4} \int\limits_0^{\tau} d\tau' \sigma_c^{\mu \nu} \left[ D_{\mu}, D_{\nu} \right] \Biggr\rbrace \, ,
\end{eqnarray}
where
\begin{eqnarray}
    && d\Gamma (\tau) \; = \; \prod\limits_{\tau'} \left( \frac{dc^* dc}{2\pi i} \right) \exp \Biggl\lbrace - c^* (\tau) \cdot c (\tau) - \int\limits_{0}^{\tau} d\tau' c^* (\tau') \cdot \dot{c} (\tau') \Biggr\rbrace \, , \\
    && \sigma_c^{\mu \nu} = c^*\, \frac{1}{2} \left[ \gamma^{\mu}, \gamma^{\nu} \right] c  \, ,
\end{eqnarray}
$c_i$ and $c^*_i$ are spinor variables of integration. In the exponents on the right-hand side of Eq.~\eqref{inttraja}, one recognizes the action for the motion of a charged particle with some electric and magnetic dipole moments in the field $\left[ D_{\mu},D_{\nu} \right]$:
\begin{eqnarray}\label{acclass}
    S_{\psi} \; = \; \int\limits_{0}^{\tau} d\tau' \left( \frac{\dot{z}^2}{2} + e_{\psi} A\!\cdot \! \dot{z} + g_{\psi} B\!\cdot \! \dot{z} - \frac{1}{4}\, \sigma_c^{\mu \nu} \left[ D_{\mu}, D_{\nu} \right] \right) \, .
\end{eqnarray}

Let us now show that the non-local terms from Eqs.~\eqref{elect} and~\eqref{magnet}, which arise whenever one rewrites $\left[ D_{\mu},D_{\nu} \right]$ in terms of physical electric and magnetic fields, contribute nothing more than an additional $2\pi N$ ($N\in \mathbb{Z}$) term to the action~\eqref{acclass}. Such additional term of course does not contribute to the dynamics of the system, since $\exp \left( -2\pi N i \right) = 1$ in the expression~\eqref{inttraja} and thus there is no change to the functional integrals~\eqref{psitake} and~\eqref{genfunc1}. 

We start by considering the second and the third terms in the action~\eqref{acclass}. As one can see from Eq.~\eqref{inttraja},  $z(0)=z(\tau)$, which allows us to transform the integral using the Stokes' theorem:
\begin{equation}
    \oint \left( e_{\psi} A + g_{\psi} B \right) \cdot  dz = \int\limits_{\Sigma_{\psi}} d \Sigma^{\mu \nu}\, \Bigl( e_{\psi} (\partial \! \wedge \! A)_{\mu \nu} + g_{\psi} (\partial \! \wedge \! B)_{\mu \nu} \Bigr)  \, ,
\end{equation}
where the integral in the right-hand side is taken over any surface $\Sigma_{\psi}$ enclosed by the loop $z(\tau)$. Next, we use Eqs.~\eqref{elect} and~\eqref{magnet} to single out the non-local terms in the integrand. The integral over these terms is given by the following expression:
\begin{equation}\label{strint}
    \int\limits_{\Sigma_{\psi}} d \Sigma^{\mu \nu}\, \left( n\! \cdot \! \partial \right)^{-1} \Bigl( e_{\psi} (n \! \wedge \! j_m)^d_{\mu \nu} - g_{\psi} (n \! \wedge \! j_e)^d_{\mu \nu} \Bigr)  \, ,
\end{equation}
where we took into account that the non-local terms featuring the currents associated to the heavy fermion $\psi$ itself cancel each other.
Let us use an antisymmetric representation for the kernel of the $(n\! \cdot \! \partial)^{-1}$ operator~\cite{Zwanziger:1968rs, PhysRevD.3.880}:
\begin{equation}\label{kernel}
    (n\! \cdot \! \partial)^{-1} (x) = \frac{1}{2} \int\limits_0^{\infty} dv\, \Bigl( \delta^4 (x-n v) - \delta^4 (x+n v) \Bigr) \, .
\end{equation}
As we are interested in this work in axion Maxwell equations, which are used to describe the behaviour of the axion in classical electromagnetic fields, we assume that the currents of light charged particles $j_e$ (and hypothetically $j_m$) creating and probing these fields in the axion detector are given by the classical expressions~\eqref{je} and~\eqref{jm}.\footnote{This assumption can in fact be lifted: the results of this section hold for the fully quantum currents as well, since one can always convert the functional integrals over the charged fermion fields into series involving integrals over point-particle trajectories~\cite{Feynman:1948ur, Feynman:1950ir, Schwinger:1951nm, Brandt:1978wc}, so that the currents in Eq.~\eqref{strint} are represented by their classical counterparts~\eqref{je} and~\eqref{jm}.} Using Eqs.~\eqref{je}, \eqref{jm} and \eqref{kernel}, we rewrite the integral Eq.~\eqref{strint} as follows:
\begin{eqnarray}\label{link}
    &&\!\!\!\!\!\!\!\!\!\!\!\!\!\!\!\!\!\! \int\limits_{\Sigma_{\psi}} d \Sigma^{\mu \nu}\, \left( n\! \cdot \! \partial \right)^{-1} \Bigl( e_{\psi} (n \! \wedge \! j_m)^d_{\mu \nu} - g_{\psi} (n \! \wedge \! j_e)^d_{\mu \nu} \Bigr) \; = \nonumber \\
    &&\!\!\!\!\!\!\!\!\!\! \sum_i \left( e_{\psi} g_i - g_{\psi} e_i \right) \int\limits_{\Sigma^d_{\psi}} d \Sigma^{d}_{\mu \nu} \int dx_i^{\nu}\, n^{\mu} \int\limits_{0}^{\infty} dv\, \Bigl( \delta^4 (x-x_i - n v) - \delta^4 (x - x_i +n v) \Bigr) \; = \nonumber \\
    && \!\!\!\!\!\!\!\!  \qquad \quad \sum_i 2\pi m_i \int\limits_{\Sigma^d_{\psi}} d \Sigma^{d}_{\mu \nu} \int dx_i^{\nu}\, n^{\mu} \int\limits_{0}^{\infty} dv\, \Bigl( \delta^4 (x-x_i - n v) - \delta^4 (x - x_i +n v) \Bigr) \, , \;\; m_i \in \mathbb{Z} \, ,
\end{eqnarray}
where in the last step, we used the DSZ quantization condition. The integral on the right-hand side of Eq.~\eqref{link} counts the number of times the trajectory of the $i$th light charged particle intersects the oriented three-surface $\Sigma_{\psi}^d\! \times \! \pm nv, \, 0\leq v < \infty$. This number always equals some integer $L_i \in \mathbb{Z}$ except for the measure zero subset of trajectories which are locally tangent to the latter three-surface. The measure zero subset does not contribute to the path integral (put another way, the trajectory of a particle can never be known with an infinite accuracy and thus one can always slightly modify the definition of the action functional, using the method outlined in Ref.~\cite{Brandt:1976hk}, so that a given trajectory is no longer tangent to $\Sigma_{\psi}^d \! \times \! \pm nv$). Therefore, we obtain the following result:
\begin{eqnarray}\label{con1}
    \int\limits_{\Sigma_{\psi}} d \Sigma^{\mu \nu}\, \Bigl( e_{\psi} (\partial \! \wedge \! A)_{\mu \nu} + g_{\psi} (\partial \! \wedge \! B)_{\mu \nu} \Bigr) \; = \; \int\limits_{\Sigma_{\psi}} d \Sigma^{\mu \nu}\, \Bigl( e_{\psi} F_{\mu \nu} + g_{\psi} F^d_{\mu \nu} \Bigr) + 2\pi N \, , \;\; N \in \mathbb{Z}\, .
\end{eqnarray}

From Eq.~\eqref{con1}, we see that the non-local parts of the second and third terms of the action~\eqref{acclass} do not contribute to the path integral~\eqref{inttraja}. Let us now show that the same holds for the last term of this action as well. We use
\begin{equation}
    \left[ D_{\mu}, D_{\nu} \right] = -i\Bigl( e_{\psi} \left( \partial \!\wedge \! A \right)_{\mu \nu} + g_{\psi} \left( \partial \!\wedge \! B \right)_{\mu \nu} \Bigr) \, ,
\end{equation}
and Eqs.~\eqref{elect}, \eqref{magnet} to single out the non-local contribution to the integrand:
\begin{eqnarray}
    &&-\frac{1}{4}\int\limits_{0}^{\tau} d\tau' \, \sigma_c^{\mu \nu} \left[ D_{\mu}, D_{\nu} \right] \; = \; \frac{i}{4} \int\limits_{0}^{\tau} d\tau' \, \sigma_c^{\mu \nu} \left( e_{\psi} F_{\mu \nu} + g_{\psi} F^d_{\mu \nu} \right) \; + \nonumber \\
    &&\quad \quad \qquad \qquad \qquad \qquad \qquad \qquad \quad \frac{i}{4} \int\limits_{0}^{\tau} d\tau' \, \sigma_c^{\mu \nu} \left( n\! \cdot \! \partial \right)^{-1} \Bigl( e_{\psi} (n \! \wedge \! j_m)^d_{\mu \nu} - g_{\psi} (n \! \wedge \! j_e)^d_{\mu \nu} \Bigr) \, .
\end{eqnarray}
Using Eqs.~\eqref{je}, \eqref{jm} and \eqref{kernel}, we obtain for the non-local term:
\begin{eqnarray}\label{nonloc}
    && \!\!\!\!\!\!\! \frac{i}{4} \int\limits_{0}^{\tau} d\tau' \, \sigma_c^{\mu \nu} \left( n\! \cdot \! \partial \right)^{-1} \Bigl( e_{\psi} (n \! \wedge \! j_m)^d_{\mu \nu} - g_{\psi} (n \! \wedge \! j_e)^d_{\mu \nu} \Bigr) \; = \nonumber \\
    && \frac{i}{4}\, \sum_i \left( e_{\psi} g_i - g_{\psi} e_i \right)  \int\limits_{0}^{\tau} d\tau' \, \sigma_{c\, \mu \nu}^d\, n^{\mu} \int dx_i^{\nu} \int\limits_{0}^{\infty} dv\, \Bigl( \delta^4 \bigl( z(\tau')-x_i - n v \bigr) - \delta^4 \bigl( z(\tau') - x_i +n v \bigr) \Bigr) \, . \qquad
\end{eqnarray}
The latter expression is non-zero only if a given trajectory $z(\tau)$ hits any of the strings $x_i \pm nv\, , \; 0\leq v < \infty$, emanating from the charged particles. The set of all such trajectories is of measure zero in the space of all the possible trajectories, over which we integrate in Eq.~\eqref{inttraja}. Thus, the non-local term~\eqref{nonloc} does not contribute to the path integral and therefore can be omitted while calculating the matrix element~\eqref{inttraja}.

\subsection{Integration over the heavy fermion intermediate state}
We have just showed that the field $\left[ D_{\mu}, D_{\nu} \right]$ entering Eq.~\eqref{inttraja} can be redefined by continuity, as the non-local string-like terms have zero contribution to the matrix element we are interested in. As discussed before, the low energy approximation ($s\ll v_a^2$) allows us to treat the latter field as constant and homogeneous, in which case the matrix element~\eqref{inttraja} can be calculated exactly using the Schwinger method~\cite{Schwinger:1951nm}. For this, we introduce the effective Hamiltonian
\begin{equation}
\mathcal{H} = \fsl{D}^2/2 = - (\fsl{p}-e_{\psi} \fsl{A}-g_{\psi} \fsl{B})^2/2 \, ,
\end{equation}
and solve the Heisenberg equations of motion in a constant field $C_{\mu \nu} \equiv \left( \left[  D_{\mu},D_{\nu} \right] \right)_c$, where the subscript $c$ means that the field $\left[  D_{\mu},D_{\nu} \right] $ is redefined by continuity. The further calculation closely follows the one performed by Schwinger in Ref.~\cite{Schwinger:1951nm}, apart from some numerical factors, and we obtain the following result:
\begin{equation}\label{matrres}
    \langle x | e^{-i\tau \Scale[0.7]{\fsl{D}}_c^2/2} | x \rangle \; = \; -\frac{i}{4 \pi^2}\, \frac{\textrm{pf}\, (C_{\alpha \beta}/2)}{\textrm{pf} \sinh{(\tau C_{\alpha \beta}/2)}} \cdot \exp \left( -\frac{i\tau }{4}\sigma_{\mu \nu} C^{\mu \nu} \right),
\end{equation}
see also Ref.~\cite{Sokolov:2021ydn}, where we considered a more general case of a non-Abelian monopole. Using the result for the matrix element Eq.~\eqref{matrres}, we can now calculate the trace and the proper time integral in Eq.~\eqref{propt}:
\begin{eqnarray}\label{effax}
    && \!\!\!\!\!\!\!\!\!\!\!\!\!\!\!\!\!\!\!\! -\text{tr}_{\gamma} \int d^4 x \; \frac{iym}{2\sqrt{2}}\, a \gamma_5 \int\limits_0^{\infty} d\tau \, \langle x | e^{-i\tau (\Scale[0.7]{\fsl{D}_c}^2+m^2)/2}  | x \rangle \; = \; - \frac{ym}{8\pi^2 \sqrt{2}}\int d^4 x \, a \int\limits_{0}^{\infty} d\tau\, e^{-i\tau m^2/2}  \times \nonumber \\[6pt]
    && \!\!\!\!\!\!\!\!\!\!\!\! \frac{\textrm{pf}\, (C_{\alpha \beta}/2)}{\textrm{pf} \sinh{(\tau C_{\alpha \beta}/2)}} \cdot \text{tr}_{\gamma}\, \gamma_5 \exp \left( -\frac{i\tau }{4}\sigma_{\mu \nu} C^{\mu \nu} \right) \; = \;  -\frac{ym}{64\pi^2 \sqrt{2}}\int d^4 x \, a\, \epsilon_{\mu \nu \lambda \rho} C^{\mu \nu } C^{\lambda \rho} \int\limits_{0}^{\infty} d\tau\, e^{-i\tau m^2/2} \; = \nonumber \\[6pt]
    && \qquad \qquad \qquad \qquad \qquad \qquad \frac{iy}{32\pi^2 \sqrt{2} m}\int d^4 x \, a\, \epsilon_{\mu \nu \lambda \rho} C^{\mu \nu } C^{\lambda \rho} \; = \; \frac{i}{16\pi^2 v_a}\int d^4 x \, a\,  C^{\mu \nu } C_{\mu \nu}^d \; ,
\end{eqnarray}
where we used the following identity which holds for any skew-symmetric four-by-four matrix:
\begin{equation}
    \text{tr}_{\gamma} \, \gamma_5 \exp \left( -\frac{i\tau }{4}\sigma_{\mu \nu} C^{\mu \nu} \right) \; = \; 4\, \textrm{pf} \sinh{(\tau C_{\alpha \beta}/2)} \, ,
\end{equation}
as well as the expression for the Pfaffian of such a matrix: $\textrm{pf}\, C_{\alpha \beta} = \epsilon_{\mu \nu \lambda \rho} C^{\mu \nu } C^{\lambda \rho}/8$. We then rewrite the result of Eq.~\eqref{effax} in terms of the four-potentials:
\begin{eqnarray}\label{efflag}
    && \frac{i}{16\pi^2 v_a}\int d^4 x \, a\,  C^{\mu \nu } C_{\mu \nu}^d \; = \; \frac{i}{16\pi^2 v_a}\int d^4 x \, a \left( \left[  D^{\mu},D^{\nu} \right] \right)_c \left( \left[  D_{\mu},D_{\nu} \right] \right)_c^d \; = \nonumber \\[6pt]
    && -\frac{i}{16\pi^2 v_a}\int d^4 x \, a\, \Bigl( e_{\psi}^2 \left( \partial \! \wedge \! A \right)^{\mu \nu}_c \left(  \partial \! \wedge \! A \right)_{c\, \mu \nu}^d + g_{\psi}^2 \left( \partial \! \wedge \! B \right)^{\mu \nu}_c \left(  \partial \! \wedge \! B \right)_{c\, \mu \nu}^d + 2\, e_{\psi} g_{\psi} \left( \partial \! \wedge \! A \right)^{\mu \nu}_c \left(  \partial \! \wedge \! B \right)_{c\, \mu \nu}^d \Bigr) \, . \qquad \quad \; \,
\end{eqnarray}

Note that in the low energy approximation, the contribution from the second exponent in Eq.~\eqref{psitake} is an analog of the Euler-Heisenberg Lagrangian, as it describes the influence of the heavy charged particle on the low energy electromagnetic field. This contribution is known to be suppressed by integer powers of the small parameter $s^2/m^{4}$~\cite{Kovalevich:1997de}. Thus, the leading order term in the effective Lagrangian stemming from the integration over the heavy fermion is given by Eq.~\eqref{efflag}. This means that in the low energy approximation, the result for the integral over $\psi$ is:
\begin{eqnarray}
    && \!\!\!\!\!\!\!\!\!\!\!\!\! \mathcal{N} \int \prod_{\alpha} \mathcal{D} \psi_{\alpha} \mathcal{D} \bar{\psi}_{\alpha} \exp \left( i\! \int \! d^4 x\, \mathcal{L}_{\psi} \right) \; = \nonumber \\[6pt] 
    && \!\!\!\!\!\!\!\!\!\!\!\!\!  \exp \left\lbrace \frac{i}{16\pi^2 v_a}\int d^4 x \, a\, \textrm{Tr}\Bigl( e_{\psi}^2 \left( \partial \! \wedge \! A \right)_c \!\! \cdot \! \left(  \partial \! \wedge \! A \right)_c^d + g_{\psi}^2 \left( \partial \! \wedge \! B \right)_c \!\!  \cdot \! \left(  \partial \! \wedge \! B \right)_c^d + 2 e_{\psi} g_{\psi} \left( \partial \! \wedge \! A \right)_c \!\!  \cdot \! \left(  \partial \! \wedge \! B \right)_c^d \Bigr) \right\rbrace \, .  \quad \; \,
\end{eqnarray}

\section{Axion Maxwell equations}\label{sec4}

\subsection{Derivation of the axion Maxwell equations}

In the previous section, we found the effective Lagrangian describing low energy interactions between axion and electromagnetic field:
\begin{equation}\label{leff}
    \mathcal{L}_{\text{a\mbox{\scriptsize{EM}}}} \; = \; \frac{1}{16\pi^2 v_a}\, a\, \textrm{Tr}\Bigl( e_{\psi}^2 \left( \partial \! \wedge \! A \right)_c \!\! \cdot \! \left(  \partial \! \wedge \! A \right)_c^d + g_{\psi}^2 \left( \partial \! \wedge \! B \right)_c \!\!  \cdot \! \left(  \partial \! \wedge \! B \right)_c^d + 2 e_{\psi} g_{\psi} \left( \partial \! \wedge \! A \right)_c \!\!  \cdot \! \left(  \partial \! \wedge \! B \right)_c^d \Bigr) \, .
\end{equation}
An important point is that in such low energy description, the non-local string-like parts should be excluded from tensors $(\partial \wedge A) $ and $(\partial \wedge B)$, as we found out previously using the full path integral formulation. In this way, the decoupling principle fails: the presence of a heavy fermion intermediate state is felt in the IR through a continuity prescription for the fields, i.e. the fermion cannot be fully integrated out. Of course, this is not unexpected as the theory is essentially non-local. For instance, the non-local string-like terms in Eqs.~\eqref{elect} and \eqref{magnet} clearly show that, no matter how heavy the charged particle is, it leaves its imprint on the long range electromagnetic field, see also the discussion at the end of sec.~\ref{qftmagn}.

The classical equations of motion are obtained by varying the full Lagrangian $\mathcal{L} = \mathcal{L}_Z + \mathcal{L}_G + \mathcal{L}_{\text{a\mbox{\scriptsize{EM}}}}$ derived from the path integral~\eqref{genfunc1}:
\begin{eqnarray}\label{eomsax}
	&& \frac{n \! \cdot \! \partial }{n^2} \left( n\! \cdot \! \partial A^{\mu} \; - \; \partial^{\mu} n \! \cdot \! A \; - \; n^{\mu} \partial \! \cdot \! A \; - \; \epsilon^{\mu}_{\,\,\, \nu \rho \sigma} n^{\nu} \partial^{\rho} B^{\sigma} \right) \; +   \nonumber \\[6pt]
	 && \qquad \qquad \qquad \qquad \qquad \frac{e_{\psi}^2}{4\pi^2 v_a}\, \partial_{\nu}\! \left\lbrace a \, (\partial \wedge A)_c^{d} \right\rbrace^{\nu \mu} \; + \; \frac{e_{\psi} g_{\psi}}{4\pi^2 v_a}\, \partial_{\nu}\! \left\lbrace a \, (\partial \wedge B)_c^{d} \right\rbrace^{\nu \mu} \; = \; {j}_e^{\, \mu} \,\, , \\[6pt]\label{eomsax2}
	&& \frac{n \! \cdot \! \partial }{n^2} \left( n\! \cdot \! \partial B^{\mu} \; - \; \partial^{\mu} n \! \cdot \! B \; - \; n^{\mu} \partial \! \cdot \! B \; - \; \epsilon^{\mu}_{\,\,\, \nu \rho \sigma} n^{\nu} \partial^{\rho} A^{\sigma} \right) \; +  \nonumber \\[6pt]
	&& \qquad \qquad \qquad \qquad \qquad \frac{g_{\psi}^2}{4\pi^2 v_a}\, \partial_{\nu}\! \left\lbrace a \, (\partial \wedge B)_c^{d} \right\rbrace^{\nu \mu} \; + \; \frac{e_{\psi} g_{\psi}}{4\pi^2 v_a}\, \partial_{\nu}\! \left\lbrace a \, (\partial \wedge A)_c^{d} \right\rbrace^{\nu \mu} \; = \; j_m^{\, \mu} \,\, , \\[6pt]\label{eomsax3}
    && \left( \partial^2 + m_a^2 \right) a \; = \; \frac{1}{16\pi^2 v_a}\, \textrm{Tr} \Bigl( e_{\psi}^2 \left( \partial \! \wedge \! A \right)_c  \!\!  \cdot \! \left(  \partial \! \wedge \! A \right)_c^d \; +  \nonumber \\[6pt]
    && \qquad \qquad \qquad \qquad \qquad \qquad \qquad \qquad g_{\psi}^2 \left( \partial \! \wedge \! B \right)_c  \!\!  \cdot \! \left(  \partial \! \wedge \! B \right)_c^d \; + \; 2\, e_{\psi} g_{\psi} \left( \partial \! \wedge \! A \right)_c  \!\!  \cdot \! \left(  \partial \! \wedge \! B \right)_c^d \Bigr) \, .
\end{eqnarray}
where we also took into account the kinetic and mass terms for the axion which were omitted in the previous equations. The first rows in the Eqs.~\eqref{eomsax} and \eqref{eomsax2} are standard for the Zwanziger theory and simply convert to the well-known expressions $\partial^{\nu} F_{\nu \mu}$ and $\partial^{\nu} F^d_{\nu \mu}$ respectively, after one takes advantage of the Eqs.~\eqref{elect}, \eqref{magnet} and the gauge-fixing conditions. Eqs.~\eqref{elect} and \eqref{magnet} also imply $\left( \partial \! \wedge \! A \right)^d_c = \left( \partial \! \wedge \! B \right)_c = F^d$ and $\left( \partial \! \wedge \! A \right)_c = -\left( \partial \! \wedge \! B \right)^d_c = F$.

It is important that the low energy approximation we used to derive the effective Lagrangian~\eqref{leff} works only for sufficiently weak fields $F$. On the other hand, as we mentioned in the end of sec.~\ref{qftmagn}, the classical field theory featuring magnetic charges is inconsistent for continuous distribution of charges and currents due to the violation of the Jacobi identity for the gauge covariant derivatives. Therefore, the classical charges must be modeled as point-like, which means that the classical electromagnetic field $F$ necessarily becomes large in the neighbourhood of each of the charges. Thus, the only way to keep the classical weak field approximation consistent is to restrict the support of the axion-dependent terms in Eqs.~\eqref{eomsax} and \eqref{eomsax2} so that it excludes the short-distance neighbourhoods of the charges. Note that such prescription is no more than a mathematical formality: the low energy experiments for which the classical Eqs.~\eqref{eomsax}-\eqref{eomsax3} are written probe only long-distance physics, anyway. Importantly, however, the latter prescription implies  $\partial^{\mu} \left( \partial \! \wedge \! A \right)^d_{c\, \mu \nu} = \partial^{\mu} \left( \partial \! \wedge \! B \right)^d_{c\, \mu \nu} = O(\sqrt{s}/v_a)$ in Eqs.~\eqref{eomsax} and \eqref{eomsax2}. As the terms featuring these derivatives are multiplied by another power of $v_a^{-1}$, we can safely omit them in our low energy approximation.

We can now rewrite Eqs.~\eqref{eomsax}-\eqref{eomsax3} in terms of the electromagnetic field strength tensor $F$:
\begin{eqnarray}\label{axMax1}
	&& \partial_{\mu} F^{\mu \nu} + \frac{e_{\psi}^2}{4\pi^2 v_a}\, \partial_{\mu} a\, F_r^{d\, \mu \nu} - \frac{e_{\psi} g_{\psi}}{4\pi^2 v_a}\, \partial_{\mu}  a\, F_r^{\mu \nu} \; = \; {j}_e^{\, \nu} \, , \\[3pt] \label{axMax2}
	&& \partial_{\mu} F^{d\, \mu \nu} - \frac{g_{\psi}^2}{4\pi^2 v_a}\, \partial_{\mu} a\, F_r^{\mu \nu} + \frac{e_{\psi} g_{\psi}}{4\pi^2 v_a}\, \partial_{\mu}  a\, F_r^{d\, \mu \nu} \; = \; j_m^{\, \nu} \, , \\[3pt] \label{axMax3}
    && \left( \partial^2 + m_a^2 \right) a \; = \; - \frac{1}{16\pi^2 v_a}\, \Bigl( \left( e_{\psi}^2 - g_{\psi}^2 \right) F_r^{\mu \nu} F_{r\, \mu \nu}^d \; - \; 2\, e_{\psi} g_{\psi} F_r^{\mu \nu} F_{r\, \mu \nu} \Bigr) \, .
\end{eqnarray}
where the subscript $r$ denotes the restriction of the support discussed in the previous paragraph. Note that the axion Maxwell equations~\eqref{axMax1} and~\eqref{axMax2}, representing physics to order $O(\sqrt{s}/v_a)$, are fully consistent with the conservation of the electric and magnetic currents to this order: $\partial_{\mu} j_e^{\mu} = \partial_{\mu} j_m^{\mu} = O(s/v_a^{2})$. As the Eqs.~\eqref{axMax1} and~\eqref{axMax2} are linear in $F$, it is convenient to decompose the electromagnetic field $F$ into the terms that are zeroth ($F_0$) and first ($F_a$) order in the small parameter $\sqrt{s}/v_a$. The zeroth order equations are then simply the usual Maxwell equations with magnetic charges, $\partial_{\mu} F_0^{\mu \nu} = j_e^{\nu}$ and $\partial_{\mu} F_0^{\mu \nu} = j_m^{\nu}$, while the first order equations describe the interaction of the external field $F_{0\, r}$ with axions:
\begin{eqnarray}\label{axext1}
    && \partial_{\mu} F_a^{\mu \nu} + \frac{e_{\psi}^2}{4\pi^2 v_a}\, \partial_{\mu} a\, F_{0\, r}^{d\, \mu \nu} - \frac{e_{\psi} g_{\psi}}{4\pi^2 v_a}\, \partial_{\mu}  a\, F_{0\, r}^{\mu \nu} \; = \; 0 \, , \\[3pt] \label{axext2}
	&& \partial_{\mu} F_a^{d\, \mu \nu} - \frac{g_{\psi}^2}{4\pi^2 v_a}\, \partial_{\mu} a\, F_{0\, r}^{\mu \nu} + \frac{e_{\psi} g_{\psi}}{4\pi^2 v_a}\, \partial_{\mu}  a\, F_{0\, r}^{d\, \mu \nu} \; = \; 0 \, .
\end{eqnarray}
Note that after we perform a similar decomposition for the axion field: $a=a_0 + a_1$, where $a_1 = O(\sqrt{s}/v_a)$, Eq.~\eqref{axMax3} decouples from Eqs.~\eqref{axext1} and~\eqref{axext2}, as the latter two equations depend only on $a_0$, which is a free field, $\left( \partial^2 + m_a^2 \right) a_0 = 0$.

If one considers a model with several heavy fermions $\psi$, it is clear that the corresponding axion Maxwell equations include the sum of the interaction terms analogous to those of the Eqs.~\eqref{axMax1}-\eqref{axMax3} with coefficients determined by the charges $(e_{\psi}, g_{\psi})$ of fermions $\psi$. If the heavy fermions carry also colour charge, which is required to solve the strong CP problem, then we have to also sum over all colour states, i.e. for each $\psi$ the coefficient gets multiplied by the dimension of the corresponding colour representation $d(C_{\psi})$. The general form of the Eqs.~\eqref{axMax1}-\eqref{axMax3} is thus:
\begin{eqnarray}\label{genaxmax1}
    && \partial_{\mu} F^{\mu \nu} + g_{a\mbox{\tiny{EE}}}\, \partial_{\mu} a\, F_r^{d\, \mu \nu} - g_{a\mbox{\tiny{EM}}}\, \partial_{\mu}  a\, F_r^{\mu \nu} \; = \; {j}_e^{\, \nu} \, , \\[12pt] \label{genaxmax2}
	&& \partial_{\mu} F^{d\, \mu \nu} - g_{a\mbox{\tiny{MM}}}\, \partial_{\mu} a\, F_r^{\mu \nu} + g_{a\mbox{\tiny{EM}}}\, \partial_{\mu}  a\, F_r^{d\, \mu \nu} \; = \; j_m^{\, \nu} \, , \\[6pt] \label{genaxmax3}
    && \left( \partial^2 + m_a^2 \right) a \; = \; - \frac{1}{4} \Bigl( \left( g_{a\mbox{\tiny{EE}}} - g_{a\mbox{\tiny{MM}}} \right) F_r^{\mu \nu} F_{r\, \mu \nu}^d \; - \; 2\, g_{a\mbox{\tiny{EM}}}\, F_r^{\mu \nu} F_{r\, \mu \nu} \Bigr) \, ,
\end{eqnarray}
where
\begin{align}\label{gEE}
	& g_{a\mbox{\tiny{EE}}} = \frac{E}{4\pi^2 v_a} \, , \quad E = \sum_{\psi} e_{\psi}^{ 2} \cdot d\! \left( C_{\psi} \right) \, , \\[5pt]\label{gMM}
	& g_{a\mbox{\tiny{MM}}} = \frac{M}{4\pi^2 v_a} \, , \quad M = \sum_{\psi} g_{\psi}^{ 2} \cdot d\! \left( C_{\psi} \right) \, , \\[5pt]\label{gEM}
	& g_{a\mbox{\tiny{EM}}} = \frac{D}{4\pi^2 v_a} \, , \quad D = \sum_{\psi} e_{\psi} g_{\psi} \cdot d\! \left( C_{\psi} \right) \, ,
\end{align}
and we introduced the anomaly coefficients $E$, $M$ and $D$~\cite{Sokolov:2021eaz}. The first order $O(\sqrt{s}/v_a)$ equations generalizing the Eqs.~\eqref{axext1} and~\eqref{axext2} are:
\begin{eqnarray}\label{genaxext1}
    && \partial_{\mu} F_a^{\mu \nu} + g_{a\mbox{\tiny{EE}}}\, \partial_{\mu} a\, F_{0\, r}^{d\, \mu \nu} - g_{a\mbox{\tiny{EM}}}\, \partial_{\mu}  a\, F_{0\, r}^{\mu \nu} \; = \; 0 \, , \\[3pt] \label{genaxext2}
	&& \partial_{\mu} F_a^{d\, \mu \nu} - g_{a\mbox{\tiny{MM}}}\, \partial_{\mu} a\, F_{0\, r}^{\mu \nu} + g_{a\mbox{\tiny{EM}}}\, \partial_{\mu}  a\, F_{0\, r}^{d\, \mu \nu} \; = \; 0 \, .
\end{eqnarray}
Rewritten in terms of the electric and magnetic fields, these equations become:
\begin{eqnarray}\label{rotbax}
    &&\pmb{\nabla}\! \times \! \mathbf{B}_a - \dot{\mathbf{E}}_a = -g_{a\mbox{\tiny{EE}}} \left( \mathbf{E}_0\! \times \! \pmb{\nabla} a - \dot{a} \mathbf{B}_0 \right) - g_{a\mbox{\tiny{EM}}} \left( \mathbf{B}_0\! \times \! \pmb{\nabla} a + \dot{a} \mathbf{E}_0 \right) \, , \\[3pt]\label{roteax}
	&&\pmb{\nabla}\! \times \! \mathbf{E}_a + \dot{\mathbf{B}}_a = g_{a\mbox{\tiny{MM}}} \left( \mathbf{B}_0\! \times \! \pmb{\nabla} a + \dot{a} \mathbf{E}_0 \right) + g_{a\mbox{\tiny{EM}}} \left( \mathbf{E}_0\! \times \! \pmb{\nabla} a - \dot{a} \mathbf{B}_0 \right) \, , \\[3pt]\label{divbax}
	&& \pmb{\nabla}\! \cdot \! \mathbf{B}_a =  g_{a\mbox{\tiny{MM}}}\, \mathbf{E}_0\! \cdot \! \pmb{\nabla} a - g_{a\mbox{\tiny{EM}}}\, \mathbf{B}_0\! \cdot \! \pmb{\nabla} a \, , \\[3pt]\label{diveax}
	&& \pmb{\nabla}\! \cdot \! \mathbf{E}_a = - g_{a\mbox{\tiny{EE}}}\, \mathbf{B}_0\! \cdot \! \pmb{\nabla} a + g_{a\mbox{\tiny{EM}}}\, \mathbf{E}_0\! \cdot \! \pmb{\nabla} a \, ,
\end{eqnarray}
where we omitted the subscript $r$ in order to conform with the conventional notation for external fields.

\subsection{Implications of the axion Maxwell equations}

The phenomenological consequences of Eqs.~\eqref{rotbax}-\eqref{diveax} were discussed in Refs.~\cite{Sokolov:2022fvs, Li:2022oel, Tobar:2022rko, McAllister:2022ibe, Thomson:2023moc, Tobar:2023rga}. Here, we would like to review the main results, without going into details.

First, let us note that Eqs.~\eqref{roteax} and~\eqref{divbax} show that there can appear effective magnetic charges and currents in the presence of axions in external electromagnetic fields. This means that Faraday's law and the no-magnetic-monopoles law can be violated in a given experiment with external $\mathbf{E}_0$ or $\mathbf{B}_0$ field assuming there exists some cosmic abundance of axion-like particles. As these laws cannot be violated in the case where no magnetic monopoles exist\footnote{This is obviously true in the flat space-time case we consider here; for the curved space-time case see~Ref.~\cite{Sokolov:2022dej}.}, one can experimentally test the existence of heavy magnetic monopoles and dyons. Such indirect probe of heavy dyons would complement the numerous experiments searching for cosmic magnetic monopoles~\cite{ParticleDataGroup:2022pth}, as it does not depend on the cosmic abundance of monopoles and dyons.

Second, note that due to the DSZ quantization condition applied to Eqs.~\eqref{gEE}-\eqref{gEM}, one expects $g_{a\mbox{\tiny{MM}}} \gg g_{a\mbox{\tiny{EM}}} \gg g_{a\mbox{\tiny{EE}}}$. Indeed, the electric charge of the electron $e \ll 1$ is small while its magnetic charge is zero, so substituting these charges into the DSZ quantization condition applied to the pair of particles -- an electron and a given heavy dyon $\psi$ -- we see that $g_{\psi} = 2\pi n / e \gg 1$ ($n \in \mathbb{Z}$) for every dyon. Then, the DSZ quantization condition applied to dyons can be solved as an equation for the electric charges $e_{\psi}$ of the dyons: $e_{\psi} = (n^e_i+\theta n^m_i / 2\pi ) \cdot e$ (where $n^e_i, n^m_i \in \mathbb{Z}$, while $\theta \in [0,2\pi )$ is the CP-violating parameter), which assuming $n^e_i, n^m_i \sim 1$ gives $e_{\psi} \sim e \ll 1$. Therefore, one can expect $e_{\psi} \ll g_{\psi}$ and thus $g_{a\mbox{\tiny{MM}}} \gg g_{a\mbox{\tiny{EM}}} \gg g_{a\mbox{\tiny{EE}}}$, as mentioned previously.
Because of this, the effective axion-induced magnetic current is expected to dominate over the effective axion-induced electric current in a given axion detection experiment, which significantly changes the response of the system to the axion dark matter signal compared to the normally considered case described by Eqs.~\eqref{axmax1}-\eqref{axmax4}. For instance, this means that in the case where the axion wavelength is much larger than the size of the detector, the dominant effect is given by the axion-induced electric field $\mathbf{E}_a$, as opposed to $\mathbf{B}_a$. This represents a clear distinction from the conventional case,  cf. Eqs.~\eqref{axmax1}-\eqref{axmax4}, where the dominant axion-induced field in the long-wavelength case is $\mathbf{B}_a$. In the case of resonant haloscope experiments, it turns out that the DC magnetic field $\mathbf{B}_0$ normally used yields a system which is not sensitive to the dominant $g_{a\mbox{\tiny{MM}}}$ and $g_{a\mbox{\tiny{EM}}}$ couplings in the non-relativistic limit $|\mathbf{k}_a|/\omega_a \to 0$. On the contrary, applying DC electric field $\mathbf{E}_0$ to a cavity resonator, one can achieve sensitivity to the latter two couplings. In general, the new couplings $g_{a\mbox{\tiny{MM}}}$ and $g_{a\mbox{\tiny{EM}}}$ provide a lot of unique signatures; the corresponding effects can be easily distinguished from the effects of the conventional $g_{a\gamma \gamma} = g_{a\mbox{\tiny{EE}}}$ coupling in a wide range of experiments.

Third, the hierarchy $g_{a\mbox{\tiny{MM}}} \gg g_{a\mbox{\tiny{EM}}} \gg g_{a\mbox{\tiny{EE}}}$ has important implications for the searches for the QCD axion, i.e. the axion which solves the strong CP problem. In this case, the axion decay constant $f_a = v_a/2N$, $2N \in \mathbb{Z}$, is fixed in terms of the axion mass $m_a$, which means that the axion couplings \eqref{gEE}-\eqref{gEM} can be plotted as functions of $m_a$, with uncertainties given by the anomaly coefficients $N$, $E$, $M$ and $D$. The hierarchy of the couplings implies that at a given mass $m_a$, the effects of the $g_{a\mbox{\tiny{MM}}}$ and $ g_{a\mbox{\tiny{EM}}}$ couplings are much stronger compared to the effects of the conventional $g_{a\gamma \gamma} = g_{a\mbox{\tiny{EE}}}$ coupling. This means that our current experiments, if adapted to look for the new couplings, can be much more sensitive to the electromagnetic couplings of the QCD axion than previously thought. Moreover, it turns out that the $g_{a\mbox{\tiny{MM}}}$ coupling of the QCD axion can explain the anomalous TeV transparency of the Universe~\cite{DeAngelis:2008sk, Horns:2012fx, Troitsky:2020rpc} -- an issue which was hypothesised to be resolved by an axion-like particle with only the $g_{a\gamma \gamma} = g_{a\mbox{\tiny{EE}}}$ coupling in many investigations, but which could not be explained in the framework of conventional QCD axion models with only the $g_{a\gamma \gamma} = g_{a\mbox{\tiny{EE}}}$ coupling, since in the KSVZ and DFSZ models, the value of the coupling $g_{a\gamma \gamma} \lesssim 10^{-16}~\text{GeV}^{-1}$ is too small in the relevant range of masses $m_a \sim 10^{-9} - 10^{-6}~\text{eV}$. Even more, the $g_{a\mbox{\tiny{MM}}}$ coupling, required to explain the anomalous TeV transparency, can also account for another astrophysical axion hint, which was derived from studying the cooling of horizontal branch stars in globular clusters~\cite{Ayala:2014pea}. Note that this latter hint cannot be explained within the KSVZ and DFSZ models assuming the mass range $m_a \sim 10^{-9} - 10^{-6}~\text{eV}$ -- the axion-photon coupling $g_{a\gamma \gamma} \lesssim 10^{-16}~\text{GeV}^{-1}$ calculated in these models is too small.

Fourth, the coupling $g_{a\mbox{\tiny{EM}}}$ violates CP. Note that the conventional coupling $g_{a\gamma \gamma} = g_{a\mbox{\tiny{EE}}}$ is necessarily CP-conserving, since in general, QED preserves CP. In the QFT with dyons, however, there exists a possible source of CP violation associated to the charge spectrum of dyons of the theory. This means that the CP violation in the electromagnetic interactions of axions is a clear signature of the existence of heavy dyons. Moreover, such CP violation would imply that the spectrum of heavy dyons is CP-violating, i.e. $D\neq 0$ in Eq.~\eqref{gEM}. Experimentally, one can probe the $g_{a\mbox{\tiny{EM}}}$ coupling in light-shining-through-wall experiments by varying the polarization of the incoming light~\cite{Sokolov:2022fvs}, as well as in haloscope experiments~\cite{Flambaum:2022zuq, Tobar:2022rko}.

Finally, let us note that the axion Maxwell equations~\eqref{rotbax}-\eqref{diveax} become trivial in the case of a constant and homogeneous axion field $a/v_a=\theta$. This means that there is no Witten-effect induced interaction~\cite{Fischler:1983sc} between the axion and the currents of charged particles $j_e$ and $j_m$ in the model we consider. In particular, charged particles do not obtain extra charges proportional to $\theta$ from their interaction with the axion field. Contrary to the misconception which sometimes appears in the literature, the Witten-effect induced interactions of axions are \textit{not} a general feature of axion electrodynamics.\footnote{The latter misconception probably originates from an unjustified extrapolation of  Eqs.~\eqref{axMax1} and \eqref{axMax2} into the domain where the weak field approximation is no longer valid, i.e. by neglecting the necessary restriction on the domain of the axion-dependent terms; see our discussion preceding  Eqs.~\eqref{axMax1} and \eqref{axMax2}. See also a strict mathematical explanation for the case of the Witten effect itself, i.e. constant $a/v_a = \theta$, in Appendix~\ref{A}.} The Witten-effect induced interactions arise only in particular ultraviolet (UV) models which feature an extra rotor (instanton) degree of freedom in the IR, see Ref.~\cite{Sokolov:2022fvs} for a detailed discussion.

\section{Summary}\label{seclast}

Using the path integral approach, we gave a detailed step-by-step derivation of the axion Maxwell equations in dyon-philic axion models, i.e. in hadronic axion models where heavy PQ-charged quarks are allowed to carry magnetic charges. The form of the derived axion Maxwell equations~\eqref{rotbax}-\eqref{diveax} fully agrees with the one derived by us in the EFT approach in a previous publication~\cite{Sokolov:2022fvs}. In particular, we confirmed that there can exist additional axion-photon couplings $g_{a\mbox{\tiny{MM}}}$ and $g_{a\mbox{\tiny{EM}}}$ along with the normally considered axion-photon coupling $g_{a\gamma \gamma} = g_{a\mbox{\tiny{EE}}}$. As these new couplings change the structure of the axion Maxwell equations significantly, we predict unique signatures in haloscope and light-shining-through-wall experiments. In particular, the detection of the effective axion-induced magnetic charges or currents in a haloscope would represent an indirect evidence for the existence of magnetically charged matter. Moreover, the new electromagnetic couplings of axions can reconcile the Peccei-Quinn solution to the strong CP problem with astrophysical axion hints, such as the anomalous TeV transparency of the Universe and the anomalous energy loss of horizontal branch stars in globular clusters.

Through an example of a particular UV-complete dyon-philic axion model, the path integral approach allowed us to illustrate and explain in detail some peculiarities of the low energy description of QFTs with magnetic charges, namely the impossibility to fully integrate out heavy dyons and the ensuing continuity prescriptions for the fields in the IR. Also, by analyzing the electromagnetic interactions of axions in the dyon-philic axion models, we found that in these models, there are no Witten-effect induced interactions between axions and charged particles. Thus, we confirmed that the axion-photon couplings and the Witten-effect induced couplings need not coincide: an important fact discussed in detail in Ref.~\cite{Sokolov:2022fvs}.

\medskip
\section*{Acknowledgements}
A.S. is funded by the UK Research and Innovation grant MR/V024566/1. A.R. acknowledges support by the Deutsche Forschungsgemeinschaft (DFG, German Research Foundation) under Germany’s Excellence Strategy - EXC 2121 Quantum Universe - 390833306. This work has been partially funded by the Deutsche Forschungsgemeinschaft (DFG, German Research Foundation) - 491245950.

\medskip

\appendix
\section{$\theta$-term in $U(1)$ gauge theories}\label{A}

One can sometimes encounter incorrect claims in the literature that the $\theta$-term in $U(1)$ gauge theories leads to the Witten effect in the presence of magnetic monopoles, i.e. it is claimed that adding the following term to the Lagrangian,
\begin{equation}\label{thetaterm}
\mathcal{L} \; \supset \; \theta F^{\mu \nu} F^d_{\mu \nu}\, ,
\end{equation}
generates electric charge proportional to $\theta$ for every magnetic monopole of the theory. Such statement would obviously contradict the result obtained by an explicit calculation in the main body of this article, since it would imply that the standard axion-photon coupling $g_{a\gamma \gamma}$ is the same as the Witten-effect induced axion-photon coupling, while we obtained a non-zero $g_{a\gamma \gamma}$ and zero Witten-effect induced coupling. Let us now remind the reader why the above-mentioned statement about the Witten effect in $U(1)$ gauge theories is incorrect. 

Mathematically, $U(1)$ gauge theory in space-time $M$ is described by a principal $U(1)$-bundle $\pi : P \rightarrow M$. The curvature 2-form $F$ of the latter principal bundle does not depend on the choice of the principal connection 1-form $iA$ valued in the Lie algebra $\mathfrak{u}(1) \cong i\mathbb{R}$. Moreover, due to the Abelian nature of the structure group, the relation between these two forms is particularly simple: $F=dA$. Given that the curvature $F\! \in \Omega^2(M)$\footnote{$\; \Omega^2(M)$ is a standard mathematical notation for the set of all differential 2-forms on $M$.} is $U(1)$-invariant, the action functional for the theory is defined on the space of all the connections $\mathcal{A}(P)$ as follows:
\begin{eqnarray}\label{u1}
    S [ A ] = \int_M \left( -\frac{1}{2e^2}\,  F \wedge \hodge F \; + \; \frac{\theta}{4\pi^2}\, F\wedge F \right) \, .
\end{eqnarray}
If one assumes $M=T\times \mathbb{R}^3$, which is topologically equivalent to the physical case of flat space-time, then the considered principal bundle is trivial, and no charge quantization arises. 

However, we are interested in studying the $U(1)$ gauge theory in the presence of a magnetic monopole, which yields a completely different scenario: in this case, the topology of $M$ has to be modified. Indeed, suppose there exists a static magnetic monopole at the origin of the spatial coordinates $\lbrace 0 \rbrace \in \mathbb{R}^3$, and consider any sphere $S^2_0$ with the centre at $\lbrace 0 \rbrace$. By definition of the monopole, the magnetic flux $\int_{S^2_0} F \neq 0$, which due to the identity $F=dA$ means that there cannot exist a global projection of the principal connection $A$ onto the sphere $S^2_0$ and therefore onto the base space $M$. The obstruction to the triviality of the principal bundle is described by the relevant characteristic class, which for the principal $U(1)$-bundle is the first Chern class $c_1 \in H^2(M, \mathbb{Z})$ valued in the second cohomology group of $M$ with integer coefficients. Since the corresponding Chern form, i.e. the invariant polynomial is simply $-F/2\pi$, and  $H^2(S^2_0, \mathbb{Z}) \cong \mathbb{Z}$, the magnetic flux is quantized: $\int_{S^2_0} F = 2\pi n$, $n \in \mathbb{Z}$, which of course yields the Dirac quantization condition for the charges. Now, we see exactly why the topology of $M$ should be modified: the second cohomology group is always non-trivial in the presence of a magnetic charge, and therefore our space $\mathbb{R}^3$ actually has a hole: $M = T\times \mathbb{R}^3 \backslash \lbrace 0 \rbrace$.

Note that due to the hole in the base space $M$, it is meaningless to assign any value to $F$ at $\lbrace 0 \rbrace$: this point is simply excluded from our space. For example, within the $U(1)$ gauge theory reviewed in the previous paragraphs, it does not make sense to write $\epsilon^{0\mu \nu \lambda}\partial_{\mu} F_{\nu \lambda} = 4\pi \delta^3 (\mathbf{x})$: the support of the distribution falls exactly into the hole invalidating the statement. Of course, one could choose a different, more complete theoretical framework, such as Dirac's classical relativistic theory of magnetic charges~\cite{Dirac:1948um} or Zwanziger's (Schwinger's) quantum relativistic theory of magnetic charges~\cite{Schwinger:1966nj, PhysRevD.3.880}, where no such holes arise; however, in the latter theories, one has to introduce the magnetic charges directly into the Lagrangian, as well as to modify the structure of the $U(1)$ gauge theory significantly, as it was illustrated in sec.~\ref{qftmagn} by the example of the Zwanziger theory. In Ref.~\cite{Sokolov:2022fvs}, we give a description of the Witten effect (as well as the Rubakov-Callan effect) within the Zwanziger theory.

Now, staying in the framework of the $U(1)$ gauge theory described earlier in this Appendix, let us derive the variation of the action~\eqref{u1} in the presence of a magnetic monopole and show that there arises no contribution from $\theta$, i.e. that no electric charge for the monopole is generated by the $\theta$-term~\eqref{thetaterm}. First, we decompose the electromagnetic field into the dynamical field $F_l$ and the background field $F_b$: $F = F_l + F_b$, where $F_b$ is the field of the monopole. As the monopole is an external source, we have to vary only the four-potential $A_l$ ($F_l = dA_l$), which is defined globally everywhere in $M = T\times \mathbb{R}^3 \backslash \lbrace 0 \rbrace$, because there are no dynamical magnetic charges by assumption. Second, we describe the boundary of the hole in $M$ by an infinitesimal sphere $S^2_{\varepsilon}$ with radius $r \to 0$, and the boundary of $M$ at infinity as a sphere $S^2_{\infty}$ with radius $r \to \infty$, as usual. The variation of the action~\eqref{u1} is:
\begin{eqnarray}\label{varacu1}
    \delta S \; = \; -\frac{1}{e^2} \int_M   d \hodge F \wedge \delta A_l \; + \; \int_T \left( \int_{S^2_{\varepsilon}} K + \int_{S^2_{\infty}} K \right) \, , \qquad K = \left( -\, \hodge F/e^2 + \theta F / 2\pi^2 \right) \wedge \delta A_l \, ,
\end{eqnarray}
where we used the Bianchi identity $dF = 0$ which is valid everywhere in $M =  T\times \mathbb{R}^3 \backslash \lbrace 0 \rbrace$\footnote{Since $F$ is an invariant polynomial corresponding to the principal $U(1)$-bundle we consider, one can also view $dF = 0$ as one of the statements of the renowned Chern-Weyl theorem, although quite trivial in this case due to $d^2 = 0$.}, and omitted the total time derivatives in the integrand. Finally, we formally put the variations at the boundaries to zero: $\delta A_l |_{S^2_{\infty}} = \delta A_l |_{S^2_{\varepsilon}} = 0$, i.e. we assume that the boundary integrals in the variation of the action~\eqref{varacu1} vanish (see a detailed justification further in the text). The resulting Euler-Lagrange equations are:
\begin{eqnarray}
    d \hodge F  = 0 \, ,
\end{eqnarray}
so we see that there is no extra electric current induced by the $\theta$-term~\eqref{thetaterm}. 

In the end of this Appendix, let us analyze whether there could arise any non-trivial effects from the surface terms in the variation of the action~\eqref{varacu1}. Although within the Euler-Lagrange method, we put the variations of fields at the boundaries to zero, it is known that in non-Abelian gauge theories, semi-classical (finite-action) IR processes are possible which yield $\delta A_n |_{S^2_{\infty}} \neq 0$ (non-zero, but a pure gauge), where $A_n$ is the non-Abelian four-potential~\cite{Belavin:1975fg}. These processes called instantons can modify the IR dynamics of the non-Abelian theory. However, no similar effects arise in the $U(1)$ gauge theory case under consideration. Indeed, due to the finiteness of the action, the field strength $F$ has to fall faster than $\rho^{-2}$ at infinity, while the four-potential falls as $\delta A_l = O(\rho^{-1})$, where $\rho = \sqrt{x_{\mu} x_{\mu}}$. This means that the integral of $K$ over the infinity 3-surface $\Sigma_{\infty}$ equals zero in the limit $\rho \to \infty$: no instanton solution exists. Also, this is clear from the topology of the $U(1)$ group, since $\pi_3 (U(1)) = 0$, and no non-trivial winding is possible. 

It remains to analyze the boundary term at $S^2_{\varepsilon}$. It may seem confusing that the condition $\delta A_l |_{S^2_{\varepsilon}} = 0$, imposed by the Euler-Lagrange method, has to be satisfied: why should the value of the four-potential be fixed in the vicinity of $\lbrace 0 \rbrace$? Actually, we will show now that the $S^2_{\varepsilon}$ surface term in the variation of the action vanishes regardless of whether we impose this condition. First, let us note that for an infinitesimal sphere, $A_l$ must be the same at every point of it: we can always take the radius $r_{\varepsilon}$ of the sphere to be much smaller than the spatial variation of the four-potential. If the monopole has structure originating from some UV physics, then this is simply a statement that the field $A_l$ is a low energy field by the definition of the IR theory, and therefore its spatial variation is negligible compared to the UV short-distance scale. A homogeneous vector field integrated over $S^2_{\varepsilon}$ gives zero, so the only non-zero contribution to the surface integral could come from the terms in the integrand containing the background $F_b$ field:
\begin{eqnarray}\label{Kcalc}
    \int_T \int_{S^2_{\varepsilon}} K \; = \; \int_T \int_{S^2_{\varepsilon}} \left( -\, \hodge F_b/e^2 + \theta F_b / 2\pi^2 \right) \wedge \delta A_l \; = \; \frac{\theta}{2\pi^2} \int_T dt\, \delta \phi_l|_{S^2_{\varepsilon}} \int_{S^2_{\varepsilon}} d\mathbf{S}_{\varepsilon}\! \cdot \! \mathbf{B}_b  \; , 
\end{eqnarray}
where $\delta \phi_l|_{S^2_{\varepsilon}}$ is the variation of the scalar potential on $S^2_{\varepsilon}$, $\mathbf{B}_b$ is the background magnetic field, and $d\mathbf{S}_{\varepsilon}$ is an oriented element of $S^2_{\varepsilon}$. The surface integral at the end of the Eq.~\eqref{Kcalc} is simply the magnetic charge of the monopole, which is time-independent by construction, so we conclude that the pure gauge variations $\delta \phi_l|_{S^2_{\varepsilon}} = \dot{\alpha}$ change the Lagrangian of the system by a total time derivative and therefore are irrelevant. This means we can always set $\delta \phi_l|_{S^2_{\varepsilon}}$ to zero by performing a suitable gauge transformation at the boundary sphere $S^2_{\varepsilon}$. The surface integral~\eqref{Kcalc} has thus no relevance for the dynamics of the system.

To conclude, we showed that the total variation of the action~\eqref{u1} does not depend on $\theta$, including any possible surface terms. This means that the saddle point of the path integral of the theory also has no $\theta$-dependence, and so the $\theta$-term~\eqref{thetaterm} has no influence on the physical processes at least at leading order. This means that there does \textit{not} exist any no-go theorem saying that in the presence of the $\theta$-term in a $U(1)$ gauge theory, there cannot exist magnetic monopoles without extra electric charge of $\theta e/2\pi$. Whereas we presented a detailed analysis in this Appendix, this conclusion could actually be easily guessed from the very beginning. Indeed, within the $U(1)$ gauge theory presented earlier, a magnetic monopole is an external source and one only studies the electromagnetic theory in its background. It is not surprising that one cannot change the properties of the external source, such as its electric charge, by adding extra terms into the Lagrangian: in the end, by definition of the external source all its degrees of freedom are independent of the dynamics of the theory and therefore of the Lagrangian. The Witten effect arising for example in the case of the 't~Hooft-Polyakov monopoles can of course easily be accounted for in this theory: one should give the magnetic monopole an electric charge proportional to $\theta e/2\pi$ by definition. If the mass of such monopole (i.e. unification scale) is high enough, no low energy dynamics can change the value of this charge as well as its motion. If one wants to incorporate the monopoles into the theory as dynamical sources, then one has to use a more complete theoretical framework, as it was done in the main body of this article while studying a particular axion model. The implementation of the Witten effect as well as the Rubakov-Callan effect into such more complete theoretical framework, where the monopoles are allowed to be dynamical, was discussed by us in Ref.~\cite{Sokolov:2022fvs}.

\medskip
\bibliography{AdPaxiomon}

\end{document}